\documentclass[twocolumn, times]{aastex631} 

\usepackage{natbib}
\usepackage{amsmath}
\usepackage{amssymb}
\usepackage{subfigure}
\usepackage{url}
\usepackage{CJK}
\usepackage{longtable}
\usepackage{graphicx}   
\usepackage{epsfig}
\usepackage{bm}
\graphicspath{{figure/}}

\newcommand{\Fig}{{Figure}}

\submitjournal{\textit{Astrophysical Journal Letters}}
\shorttitle{Simulation of solar homologous eruptions}
\shortauthors{Bian et al.}

\begin{document}

\title{Homologous Coronal Mass Ejections Caused by Recurring Formation and Disruption of Current Sheet within a Sheared Magnetic Arcade}

\correspondingauthor{Chaowei Jiang} \email{chaowei@hit.edu.cn}

\author[0000-0001-9189-1846]{Xinkai Bian}
\affiliation{Institute of Space Science and Applied Technology, Harbin
	Institute of Technology, Shenzhen 518055, China}

\author[0000-0002-7018-6862]{Chaowei Jiang}
\affiliation{Institute of Space Science and Applied Technology, Harbin
  Institute of Technology, Shenzhen 518055, China}

\author[0000-0001-8605-2159]{Xueshang Feng}
\affiliation{Institute of Space Science and Applied Technology, Harbin Institute of Technology, Shenzhen 518055, China}

\author[0000-0003-4711-0306]{Pingbing Zuo}
\affiliation{Institute of Space Science and Applied Technology, Harbin
	Institute of Technology, Shenzhen 518055, China}

\author[0000-0002-7094-9794]{Yi Wang}
\affiliation{Institute of Space Science and Applied Technology, Harbin Institute of Technology, Shenzhen 518055, China}

\begin{abstract}
The Sun often produces coronal mass ejections with similar structure repeatedly from the same source region, and how these homologous eruptions are initiated remains an open question. Here, by using a new magnetohydrodynamic simulation, we show that homologous solar eruptions can be efficiently produced by recurring formation and disruption of coronal current sheet as driven by continuously shearing of the same polarity inversion line within a single bipolar configuration. These eruptions are initiated by the same mechanism, in which an internal current sheet forms slowly in a gradually sheared bipolar field and reconnection of the current sheet triggers and drives the eruption. Each of the eruptions does not release all the free energy but with a large amount left in the post-flare arcade below the erupting flux rope. Thus, a new current sheet can be more easily formed by further shearing of the post-flare arcade than by shearing a potential field arcade, and this is favorable for producing the next eruption. Furthermore, it is found that the new eruption is stronger since the newly formed current sheet has a larger current density and a lower height. In addition, our results also indicate the existence of a magnetic energy threshold for a given flux distribution, and eruption occurs once this threshold is approached.
\end{abstract}

\keywords{Sun: flares; Sun: coronal mass ejections (CMEs); Sun: Magnetic fields; Methods: numerical; Sun: corona; Magnetohydrodynamic (MHD)}

\section{Introduction}
\label{sec:intro}
Coronal mass ejections (CMEs) are the most violent eruptions on the Sun. They are manifestations of a sudden release of free energy stored in the coronal magnetic fields which otherwise undergo a quasi-static evolution as driven by photospheric motions. Most of these eruptions are produced by solar active regions (ARs), in particular, from sheared magnetic polarity inversion lines (PILs) of strong field regions. Furthermore, during their lifetime on the solar disk, some flare-productive ARs can successively generate multiple eruptions from the same PIL. These eruptions have relatively short time intervals (of a few hours to at most one day) between each other, and meanwhile, they have very similar structure, as manifested in the flare emissions and CME morphology. Such kind of event has been known as homologous eruptions~\citep{zhang_are_2002}. Observations show evolution of the source regions of homologous events is often characterized by continuously shearing motion~\citep{li_sequential_2010,Romano2015, romano_homologous_2018}, sunspot rotation~\citep{regnier_evolution_2006, zhang_relationship_2008}, and flux emergence~\citep{nitta_recurrent_2001, sterling_internal_2001, ranns_emerging_2000, dun_evolution_2007, xu_homologous_2017}, all of which can inject free magnetic energy into the corona, but the overall magnetic field configuration does not change much. Therefore, the homologous eruptions should have similar initiation mechanism between each other.

Although a number of theories have been proposed to account for solar eruption~\citep{Forbes2006, Shibata2011, ChenP2011, Schmieder2013, Aulanier2014, Janvier2015}, there are only two scenarios that have been established by magnetohydrodynamic (MHD) simulations for explaining homologous eruptions, namely the breakout model~\citep{devore_homologous_2008} and the flux emergence model~\citep{chatterjee_simulation_2013}. The breakout mechanism relies on a multipolar magnetic topology with a magnetic null point situating above a core flux that is sheared by photospheric motion. An eruption is triggered by magnetic reconnection at the null point which removes the overlying restraining field of the sheared core, and after the eruption the null-point topology is restored~\citep{Antiochos1999, Lynch2008, Wyper2017}. 
\citet{devore_homologous_2008} showed that, by imposing a continual shearing motion, such mechanism can recur to produce homologous eruptions, which are, however, confined without producing CMEs. \citet{pariat_three-dimensional_2010} also found homologous behavior in the simulation of breakout eruptions, but they focused on relatively small-scale eruptions, i.e., coronal jets. Magnetic flux emergence, in which new flux rises from the interior of the Sun to the atmosphere, is also believed to be an important mechanism to initiate eruptions~\citep{Chen2000, archontis_flux_2008, fan_eruption_2010}. \citet{chatterjee_simulation_2013} realized homologous CMEs by a kinetic emergence of a highly twisted flux tube into the solar corona through imposing proper boundary conditions. In their simulation, magnetic flux rope formed and partially erupted, repeatedly, due to kink instability~\citep{Torok2005, Fan2007}. However, the time interval between two successive eruptions seems to be too short for the post-eruption field of the first one to relax and restore an equilibrium before the second one, possibly due to the too fast injection of flux at the bottom surface. 

In this Letter, we present a new MHD simulation demonstrating a simple and efficient scenario for homologous CMEs, in which the successive eruptions can be produced by recurring formation and disruption of current sheet (CS) as driven by continuously shearing of the same PIL within a single bipolar configuration. All these eruptions are initiated by the same fundamental mechanism (as established recently by an ultra-high accuracy MHD simulation, \citet{jiang_fundamental_2021}): through surface shearing motion along the PIL, a vertical CS forms quasi-statically within the sheared arcade, and once the CS is sufficiently thin such that ideal MHD is broken down, reconnection sets in and instantly triggers and further drives the eruption. The present simulation shows that each eruption does not release all the free energy but with a large amount left in the post-flare arcade below the erupting flux rope. Thus, new CS can be more easily formed by a further shearing of these arcade than by shearing a potential field arcade, which is favorable for initiating the next eruption. Furthermore, by comparing the first eruption and the subsequent eruptions, we find that the latter ones are somewhat stronger because the newly formed CS is stronger, namely, with a higher current density and a lower height.

\section{Numerical Model}
\label{sec:model}
\begin{figure}[tbp]
	\centering
	\includegraphics[width=0.5\textwidth]{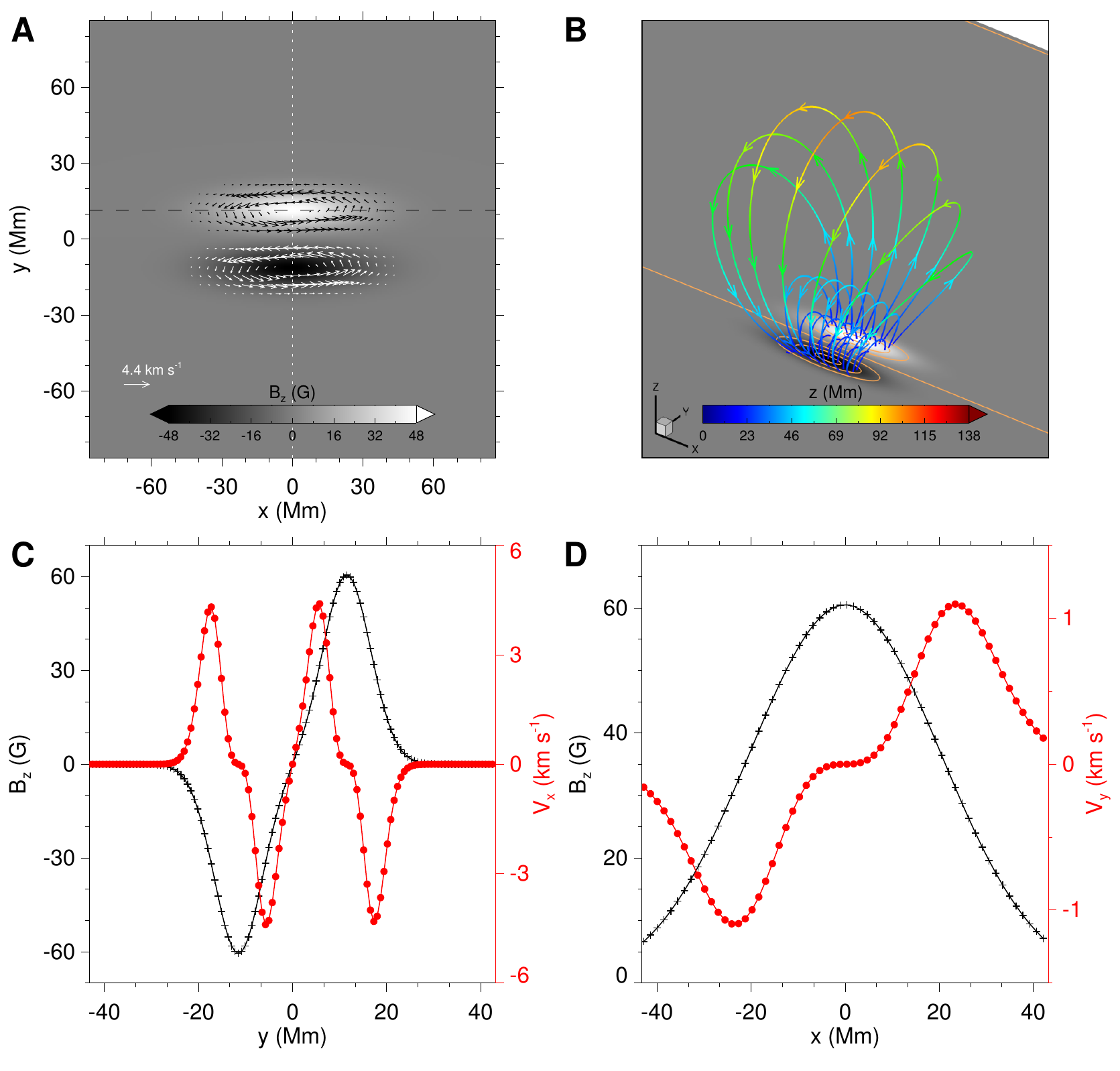}
	\caption{The initial conditions of the simulation. \textbf{A}, Magnetic flux distribution and surface flow (shown by the arrows) at the bottom surface. \textbf{B}, The 3D prospective view of the field lines of the initial potential field. The colors denote the value of the height. \textbf{C}, 
		Profile of magnetic vertical component $B_{z}$ and velocity along $(x,z)=0$ line, i.e., the white dotted line in \textbf{A}. \textbf{D}, same as \textbf{C}, but along the $(y=y|_{B_{z,\rm max}},z=0) $ line, i.e., the black dashed line in \textbf{A}.}
	\label{fig:initial cond}
\end{figure}

Our numerical model essentially follows the dynamic evolution of a bipolar coronal field as driven by continual shearing motion along its PIL at the bottom surface (i.e., photosphere), similar to our previous works~\citep{jiang_fundamental_2021, bian_numerical_2021}. To this end, we first defined a symmetrical bipolar flux distribution at the bottom surface (\Fig~\ref{fig:initial cond}A) and calculate a potential field from this flux distribution (\Fig~\ref{fig:initial cond}B). Then, an MHD simulation is started from this potential field with a background atmosphere in hydrostatic equilibrium in which the plasma is set with typical coronal values with sound speed of $110$ km s$^{-1}$, the largest Alfv$\acute{\text{e}}$n speed of $3600$ km s$^{-1}$ and the lowest plasma $\beta$ of $1.8\times10^{-3}$. The continual shearing motion is mimicked by a counterclockwise rotation flow applied to each of the two magnetic polarities. Note that the rotational flow is defined such that it will not modify the flux distribution at the bottom surface (thus no flux emergence), and more importantly, it is small enough (of a few km~s$^{-1}$, less than the local Alfv$\acute{\text{e}}$n speed by three orders of magnitude) to be a quasi-static driving of the coronal field (\Fig~\ref{fig:initial cond}C and D). More details for the definitions of the bipolar flux distribution and the surface velocity can be found in \citet{jiang_fundamental_2021} and \citet{bian_numerical_2021}, and specifically, the settings are identical to those for the simulation CASE I in \citet{bian_numerical_2021}.

The full MHD equations with both plasma pressure and solar gravity are solved in a 3D Cartesian geometry, and the numerical scheme is an advanced conservation element and solution element (CESE) method implemented on an adaptive mesh refinement (AMR) grid with parallel computing, namely, the AMR--CESE--MHD~\citep{Jiang2010, Feng2010}. Note that no explicit resistivity is used in the magnetic induction equation, but magnetic reconnection is still allowed through numerical diffusion when a current layer is sufficiently narrow such that its width is close to the grid resolution~\citep{jiang_fundamental_2021}. By this, we achieved a resistivity as small as we can with a given grid resolution. 

The computational volume spans $[-270,270]$~Mm in $x$ direction, $[-270,270]$~Mm in $y$ direction, and  $[0,540]$~Mm in $z$ direction. The full volume is resolved by a block-structured grid with AMR in which the base resolution is $\Delta x=\Delta y=\Delta z=\Delta=2.88$~Mm, and the highest resolution is $\Delta =360$~km which is used to capture the formation process of the CS and the reconnection. At the bottom surface (i.e., $z=0$), we directly solved the magnetic induction equation to self-consistently update the magnetic field as driven by the surface flow. Since our model produces multiple eruptions that will successively eject out of the computational volume, we used two types of boundary conditions according to the evolution phase, which can avoid artificial reflection of the eruption at the side and top surfaces and  maintain the long-time run of the simulation. Before the onset of the first eruption, we fixed the plasma density, temperature and velocity. The tangential components of the magnetic field are linearly extrapolated from the inner points, and the normal component is modified according to the divergence-free condition to clean numerical magnetic divergence near the boundaries. Then in the subsequent simulation process, the plasma density, temperature, velocity, and the tangential components of the magnetic field are given by the zero-gradient extrapolation from the inner points, and meanwhile the normal component of the magnetic field is also modified according to the divergence-free condition. This allows the outflowing plasma and magnetic field in the volume to pass freely out of the boundary without reflection during the eruption.

\section{Results}
\label{sec:res}
\begin{figure}[htbp]
	\centering
	\includegraphics[width=0.5\textwidth]{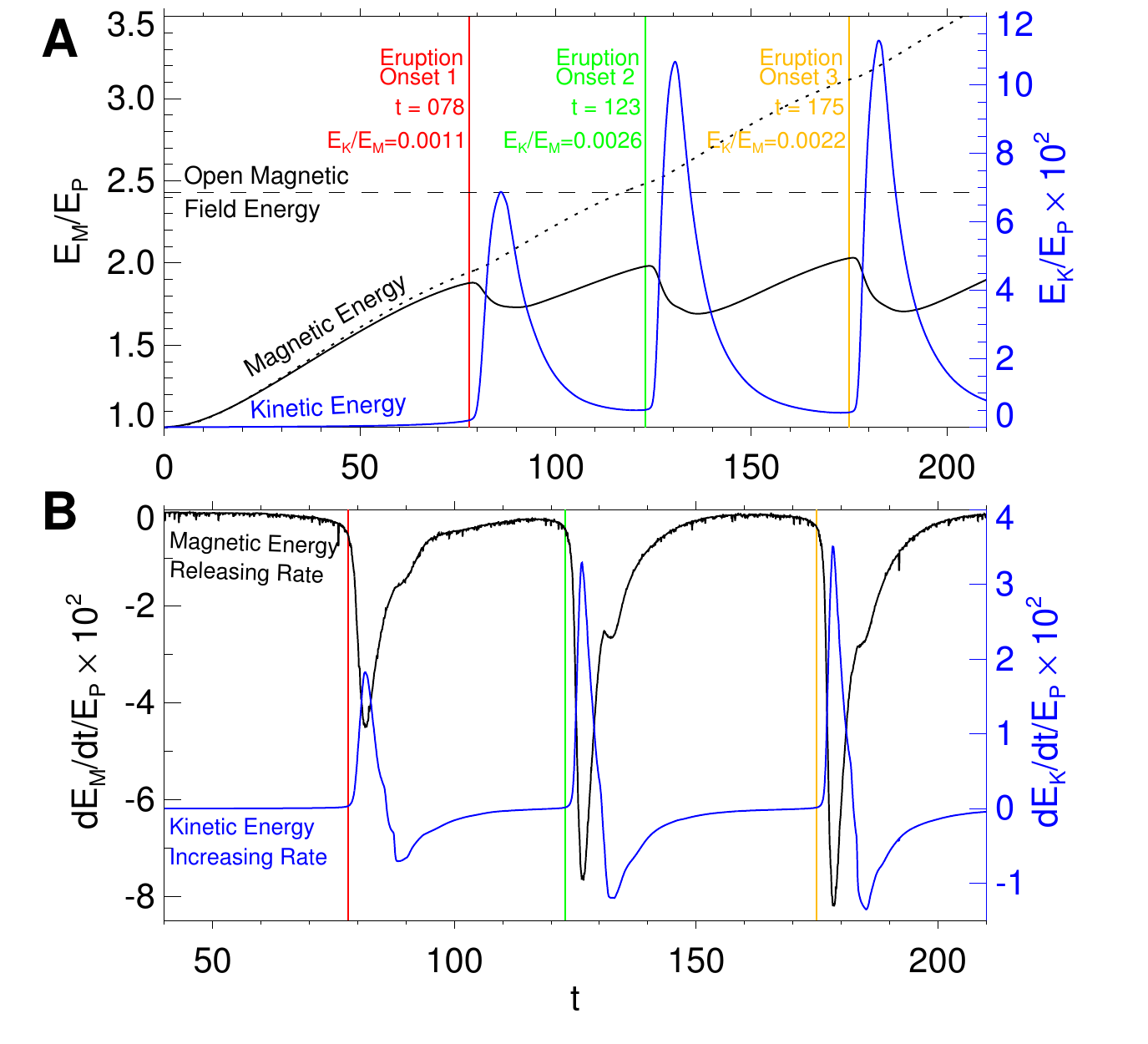}
	\caption{Temporal evolution of magnetic and kinetic energies in the simulation. \textbf{A}, Evolution of magnetic energy $E_M$ (black line) and kinetic energy $E_K$ (blue line). The dotted curve shows the energy injected into the volume (that is, a time integration of total Poynting flux) from the bottom boundary through the surface flow. The horizontal dashed line denoted the value of the total magnetic energy for an open force-free field with the same magnetic flux distribution on the bottom surface. \textbf{B}, Releasing rate of magnetic energy (black line) and the increasing rate of kinetic energy (blue line). In all the panels, the red, green, and orange lines denote the onset time of three different eruptions, respectively.}
	\label{fig:energy_evol}
\end{figure}

\Fig~\ref{fig:energy_evol} presents the evolution of the magnetic and kinetic energies during the simulation run from the initial time to $t=210$, with a time unit of $\tau = 105$ s (all the times mentioned in this paper are expressed with the same time unit). It shows that three homologous solar eruptions are produced almost in a periodic manner, with each of the eruption characterized by an impulsive increase of the kinetic energy and corresponding fast release of the magnetic energy.

\begin{figure*}[htbp]
	\centering
	\includegraphics[width=0.8\textwidth]{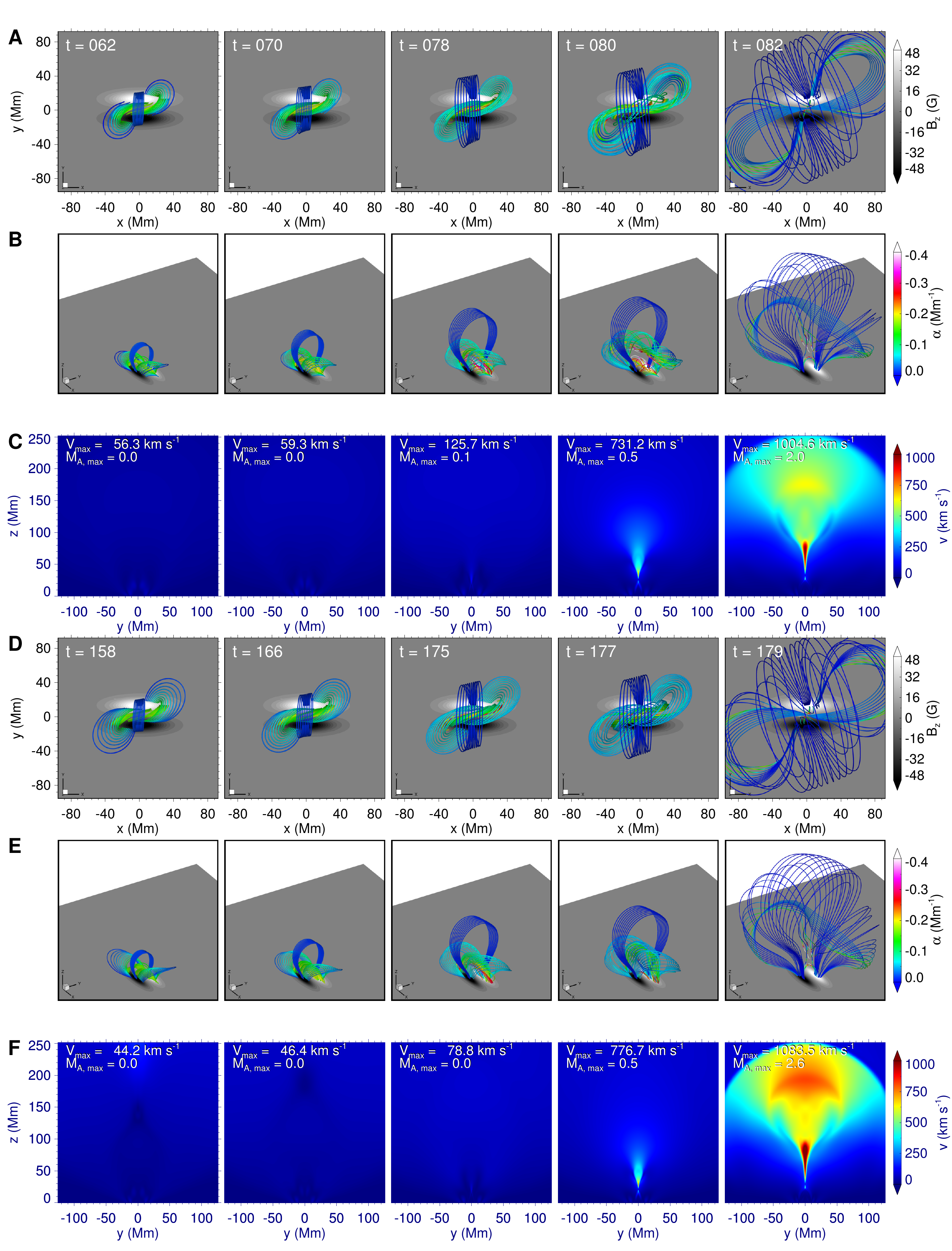}
	\caption{Evolution of magnetic field lines and velocity during the first and third eruptions of the entire simulation process. \textbf{A}, Top view of magnetic field lines. The colored thick lines represent magnetic field line and the colors denote the value of nonlinear force-free factor defined as $\alpha=\textbf{J} \cdot \textbf{B}/B^{2}$, which indicates how much the field lines are non-potential. \textbf{B}, 3D prospective view of the same field lines shown in panel \textbf{A}. \textbf{C}, Distribution of velocity on the central cross section, i.e., the $x=0$ slice. The largest velocity and Alfv$\acute{\text{e}}$nic Mach number are also denoted. \textbf{D}, \textbf{E} and \textbf{F} are the same as \textbf{A}, \textbf{B} and \textbf{C}, respectively, but at different times. \textbf{A}, \textbf{B} and \textbf{C} are the first eruption, and \textbf{D}, \textbf{E} and \textbf{F} are the third. In first two panels of \textbf{F}, the largest velocity and Alfv$\acute{\text{e}}$nic Mach number are only denoted below the reconnection region, i.e., below $100$ Mm of first panel and $150$ Mm of second panel.}
	\label{fig:mag_line}
\end{figure*}

\begin{figure*}[htbp]
	\centering
	\includegraphics[width=0.8\textwidth]{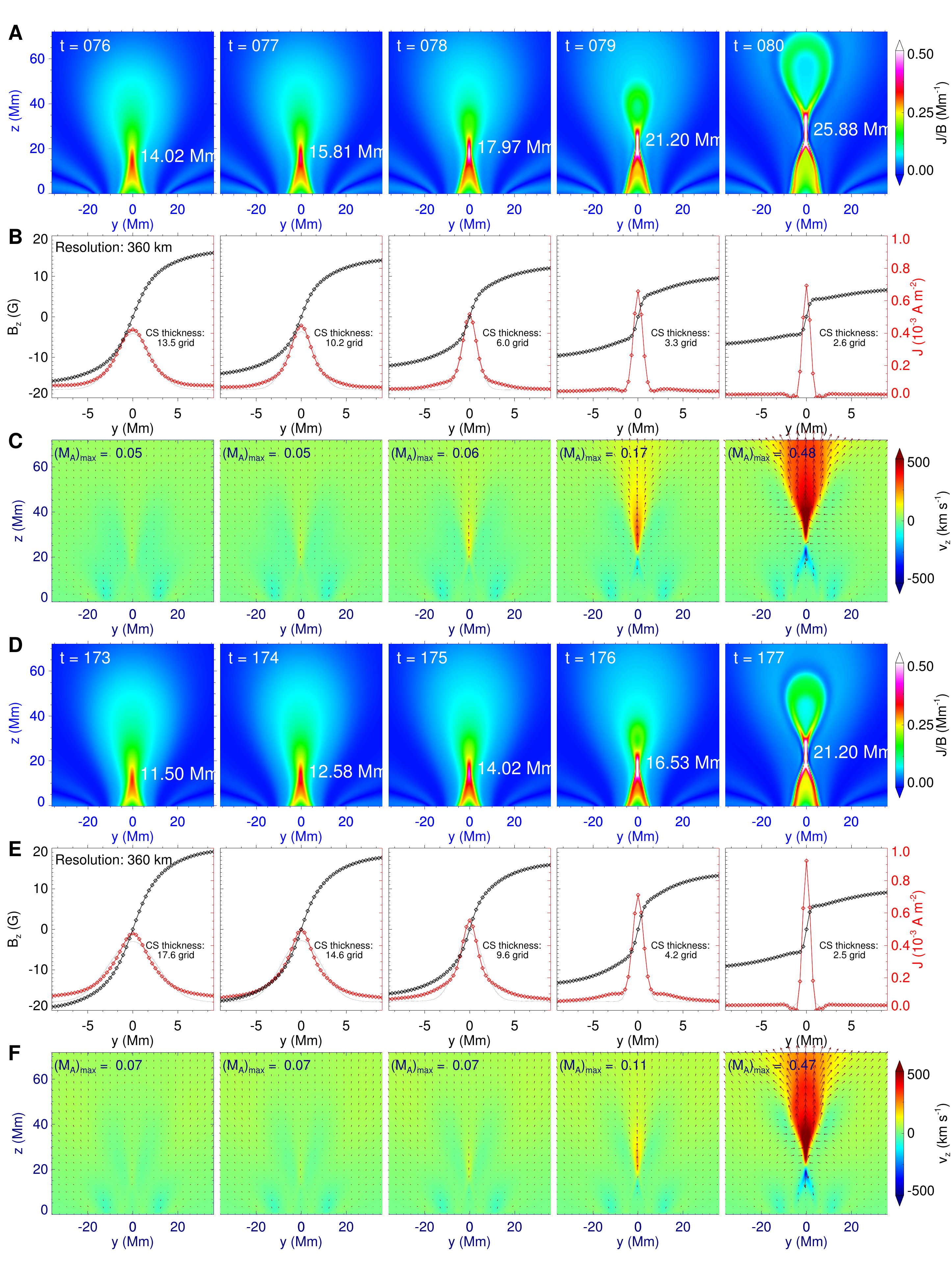}
	\caption{Formation of the CS and trigger of reconnection in the first and third eruptions. \textbf{A}, Current density \textbf{J} normalized by magnetic field strength \textbf{B} on the central vertical slice, the $x=0$ slice, at different times in the simulation. The height of maximum current density are denoted on each panel. \textbf{B}, One-dimensional profile of the magnetic field component $B_z$ and current density $J$ along a horizontal line crossing perpendicular to the CS center (that is, the point with the largest $J/B$). The diamonds denotes values on the grid nodes. The thickness of CS is denoted, which is defined by the FWHM of a Gauss function fitting (the thin black curve) of the profile of current density. \textbf{C}, Distribution of vertical velocity on the same cross-section shown in \textbf{A}. The magnitudes of maximal Alfv$\acute{\text{e}}$nic Mach number are also given. \textbf{D}, \textbf{E}, and \textbf{F} are the same as \textbf{A}, \textbf{B}, and \textbf{C}, respectively, but at different times. \textbf{A}, \textbf{B}, and \textbf{C} are the first eruption, and \textbf{D}, \textbf{E}, and \textbf{F} are the third.}
	\label{fig:JB_core}
\end{figure*}

In the early phase ($t<=76$), the shearing flow at the bottom surface continuously injects Poynting flux into the volume at nearly a constant rate (see the dashed line in \Fig~\ref{fig:energy_evol}A), so the magnetic energy increases almost linearly. In this duration, the kinetic energy is negligible (only about $10^{-3}$ of the initial magnetic energy), which indicates that the system is in a quasi-static evolution. Meanwhile, the magnetic field has evolved from the initial potential field to a configuration consisting of a strongly sheared core above the PIL that has intensive currents and an overlying envelope field that is still close to current-free (\Fig~\ref{fig:mag_line}A and B). With the increase of magnetic pressure, the core field inflates slowly (\Fig~\ref{fig:mag_line}B) and a vertical CS gradually forms above the PIL (\Fig~\ref{fig:JB_core}A and B).

At a critical time of $t=78 \sim 79$, when the thickness of the CS decreases down to the grid resolution (around $3$ grids, see \Fig~\ref{fig:JB_core}A and B), magnetic reconnection sets in and trigger an eruption. The magnetic energy drops immediately, despite the continual injection of the Poynting flux through the bottom surface. Meanwhile the kinetic energy increases impulsively to a value of nearly $7\%$ of the potential magnetic energy in a short time interval of $\sim 5\tau$. The beginning of the eruption can be seen more clearly from the time profiles of the magnetic energy release rate and the kinetic energy increase rate, both of which have a sharp increase at exactly the same time of onset of the eruption (or the reconnection). The reconnection generates a magnetic flux rope from the tip of the CS, which rises rapidly with a speed of around $600$ km s$^{-1}$ (\Fig~\ref{fig:mag_line}B and C) and eventually leaves the computational volume from the upper boundary. With ongoing of the flare reconnection, the CS arises higher and higher (see Supplementary Movie) and thus reconnection becomes weaker and weaker.

As the explosive structure gradually flows out of the computational volume, the kinetic energy of the system also gradually decreases, and returns to a very low value around $t=100$, which can also be seen from the velocity distribution (see Supplementary Movie), except that there is a slowly rising structure at the wake of the eruption. As the magnetic reconnection weakens, the magnetic energy released rate gradually decreases. When it becomes equal to the energy injection rate from the bottom surface (at around $t=90$), the total magnetic energy reaches a local minimum and then increases again. It is worth noting that the eruption releases about one third (rather than all) of the pre-eruptive free energy, with the remaining two thirds stored in the lower post-flare magnetic arcade near the PIL. This means that the post-eruption field still possesses a relatively strong magnetic shear, which provides a favorable stage for a new eruption. Moreover the post-eruption state has relaxed to an MHD equilibrium with the ratio of kinetic energy to the potential energy decreased to a low level of $10^{-3}$, close to its pre-eruptive value.
 
The second eruption follows shortly after the first eruption leaves the computational box, and the same process repeats for the third eruption. All the eruptions are initiated in the same way as a result of recurring formation of the CS in core field as driven by the shearing flow and triggering of reconnection in the CS. For example, \Fig~\ref{fig:mag_line}D to F show the pre-eruptive to eruptive process for the third eruption event, which looks very similar to the first one as shown in \Fig~\ref{fig:mag_line}A to C. Before the initiation of the eruption, the core field has relaxed to a quasi-static state from the preceding eruption, as can be seen from the velocity distribution in \Fig~\ref{fig:mag_line}F. Then the current density distribution is gradually compressed to form a vertical, narrow layer extending above the PIL, i.e., a vertical CS forms again (\Fig~\ref{fig:JB_core}D). When the thickness of the newly-formed CS drops to grid resolution, magnetic reconnection kicks in again and immediately triggers an eruption. The Supplementary Movie shows the whole evolution of the magnetic field lines, the current density, and the velocity in the successive three eruptions, all of which have similar structure among each other.

\begin{figure}[htbp]
	\centering
	\includegraphics[width=0.5\textwidth]{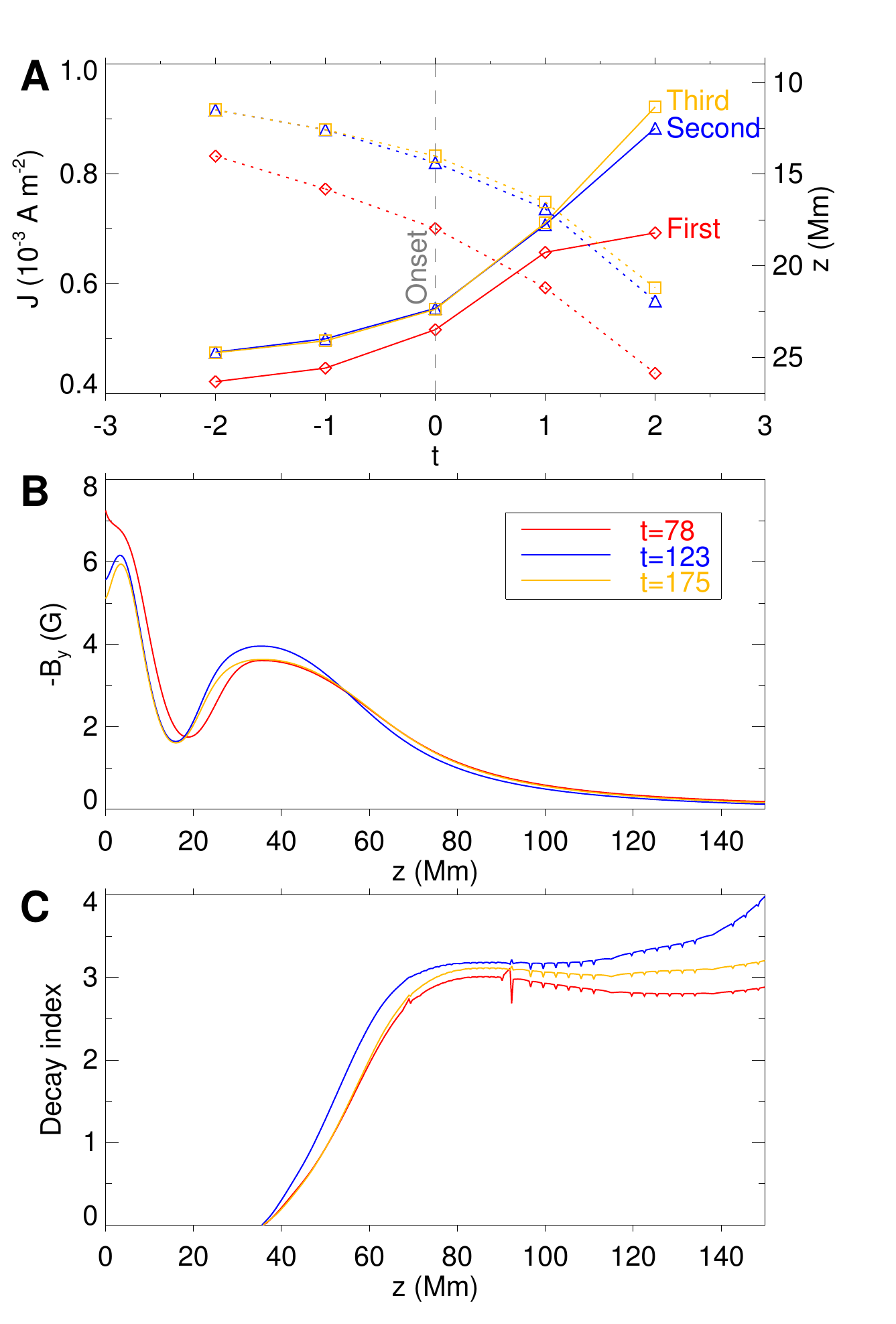}
	\caption{The CS intensity, magnetic field and decay index around the onset time of the three eruptions. \textbf{A}, Evolution of maximum current density $J$ (solid) on the $x=0$ slice and its height (dotted) with time. Note that here time $t=0$ represents the onset time of three eruptions, that is, $t=78$ for the first, $t=123$ for the second and $t=175$ for the third. Red, blue, and gold represent the first, second and third eruptions, respectively. \textbf{B} and \textbf{C} are shown for magnetic field component $B_{y}$, and decay index of $B_{y}$, respectively, along $z$ axis at the onset of three eruptions.}
	\label{fig:erupiton_intensity}
\end{figure}

From the point of view of energy, our simulation shows that the magnetic energy has an critical threshold of approximately $E_{M}=2E_{P}$ as driven by continual shear, and once it approaches this threshold, eruption will occur and release a part of the free energy, making the energy evolution a saw-toothed like profile. We note that this energy threshold is smaller than the full open field energy ($2.4 E_{P}$ for the given flux distribution), which is the energy upper limit for all force-free and simply-connected fields with a fixed flux distribution as conjectured by Aly and Sturrock~\citep{aly_how_1991, sturrock_maximum_1991}. The three eruptions are initiated at $t=78$, $123$, and $175$, respectively, with an approximately equal time cadence of around $50$. Note that such cadence is much shorter than the time from the initial time to the first eruption (namely, $78$), because each eruption only releases one third of the free energy, and thus the time for the further energy input to reach its threshold can be reduced. As can be seen from the extreme values of kinetic energy (and its increasing rate as well as the magnetic energy releasing rate), the three eruptions have small difference in their intensities of eruption, namely, the later one is somewhat stronger than its preceding one. There is an important reason for this. Our previous work~\citep{bian_numerical_2021} shows that the intensity of eruption is closely correlated to the formation height of the CS and its peak current density at the onset of the eruption. That is, the lower height CS forms at and the higher its peak current density reaches, the more efficient the reconnection is and the stronger the eruption. Comparing the CS at the onset time of these three eruptions as shown in \Fig~\ref{fig:erupiton_intensity}A, we find this, as expected, the maximum current density (the formation height) at $t=123$ and $t=175$ is slightly larger (lower) than that at $t=78$. 

We also check the influence of the overlying field on the intensity of eruption by calculating the decay index of the overlying field at the onset of these three eruptions. \Fig~\ref{fig:erupiton_intensity}B shows the overlying field $B_{y}$, which is thought to play a key role in strapping the erupting flux rope, and \Fig~\ref{fig:erupiton_intensity}C shows the decay index which is defined as $n=-{\rm d~}{\rm ln}(-B_{y})/{\rm d~}{\rm ln}(z)$. The external field and decay index of these three eruptions are basically the same, so they do not contribute to the difference in intensity of eruption.

\section{Conclusions}
\label{sec:con}
Our simulation offers a simple and efficient scenario for homologous CMEs. Unlike the existing models that often require additional conditions, such as a magnetic null point or continuously magnetic flux emergence, ours is built upon a single magnetic arcade that is stressed continuously by the photospheric shearing motion. This model is an extension of a fundamental mechanism of solar eruption~\citep{jiang_fundamental_2021}, namely, the BASIC mechanism (as called in \citet{bian_numerical_2021}) which refers to the well-established scenario that an internal CS can form slowly in a gradually sheared bipolar field and the reconnection of the CS triggers and drives the eruption. Here we show that, with the surface shearing continued, a new CS can form again above the same PIL after the completion of a previous eruption, and thus results in a new eruption. Therefore, the recurring formation and disruption of coronal CS naturally produces homologous eruptions.

In our model the field can restore to a quasi-static evolution in the interval between two eruptions, which is in agreement with observations. Moreover, being consistent with observations that post-flare loops often possess certain degree of magnetic shear, in our simulation each eruption does not release all the free energy but with a large amount left in the post-flare arcade below the erupting flux rope. Thus, CS can be more easily formed by a further shearing of these already well-sheared arcade than by shearing a potential field arcade, which is favorable for the next eruption. The evolution curve of magnetic energy forms a ``saw-tooth" shape, and the tip of each saw-tooth corresponds to the onset of an eruption (similar curve of magnetic energy evolution was also obtained in \citet{pariat_three-dimensional_2010}'s simulation for homologous coronal jets), which supports that there is an energy upper limit for the bipolar field in AR's core, that is, the open field energy corresponding to the magnetogram~\citep{aly_how_1991, sturrock_maximum_1991}. Furthermore, through the analysis of the CS and the external field at the onset of the three eruptions, it is found that the new eruption is stronger since the newly formed CS has a larger current density and a lower height, rather than the effects of external field and decay index.

In summary, our study demonstrates that homologous CMEs can be produced by recurring formation and disruption of coronal CS as driven by continuously shearing of the same PIL within a single bipolar configuration. It also indicates that there is an upper limit for energy of the bipolar field in the core of ARs, and eruption is an efficient way to keep the energy below this limit. Future studies will investigate how this mechanism works in realistic homologous events using data-driven modeling~\citep{Jiang2016NC, jiang_mhd_2021}.

\begin{acknowledgments}
\modulolinenumbers[10]
This work is jointly supported by National Natural Science Foundation of China (NSFC 42174200, 41822404, 41731067), the Fundamental Research Funds for the Central Universities (Grant No. HIT.OCEF.2021033), and Shenzhen Technology Project JCYJ20190806142609035. The computational work was carried out on TianHe-1(A), National Supercomputer
Center in Tianjin, China.
\end{acknowledgments}

\bibliographystyle{aasjournal}
\bibliography{all}

\clearpage


\end{document}